\RequirePackage{ifpdf}
\ifpdf 
\documentclass[pdftex]{sigma}
\else
\documentclass{sigma}
\fi

\newcommand{\R}{\mathbb{R}}
\newcommand{\C}{\mathbb{C}}

\newcommand{\id}{\mathbb{I}}

\newcommand{\mone}{^{-1}}

\def\dag{^\dagger}

\def\ka{\kappa}
\def\cc{{\cal C}}
\def\la{\langle}
\def\ra{\rangle}
\def\ot{{\,\otimes \,}}

\def\com#1{[ #1 ]}
\def\mn{{\mu\nu}}

\def\g{\mathfrak{g}}
\def\demi{\frac{1}{2} }
\def\dr{{\rightarrow}}
\newcommand{\one}{\mbox{$1 \hspace{-1.0mm}  {\bf l}$}}
\def\cop{{\bigtriangleup}}

 \def\bb{{\cal B}} \def\cc{{\cal C}} \def\dd{{\cal D}}   \def\ggg{{\cal G}}
\def\hh{{\cal H}}      \def\nn{{\nonumber}}
 \def\ppp{{\cal P}}   \def\sss{{\cal S}} \def\tt{{\cal T}} 
 \def\ww{{\cal W}}

\def\mmm{{\mathfrak{m}}}
\def\ppp{{\mathfrak{p}}}
\def\ss{{\mathfrak{s}}}
\def\et{{\frac{\eta}{2}}}
\def\etaa{{\frac{\eta_1}{2}}}
\def\etab{{\frac{\eta_2}{2}}}
\def\alp{{\frac{\alpha}{2}}}
\def\thet{{\frac{\theta}{2}}}

\def\hphi{{\hat \phi}}

\def\bicross{{\blacktriangleright\hspace{-1.75mm}\lhd }}

\def\act{{\, \triangleright\, }}

\def\double{{\triangleright\hspace{-0.5mm}\triangleleft}}

\def\i{{\textrm{id}}}

\newcommand{\an}{\mathfrak{an}}
\newcommand{\so}{\mathfrak{so}}

\newcommand{\AN}{\mathrm{AN}}

\newcommand{\SO}{\mathrm{SO}}
\newcommand{\SL}{\mathrm{SL}}
\newcommand{\sll}{\mathfrak{sl}}

\numberwithin{equation}{section}

\begin{document}

\allowdisplaybreaks

\renewcommand{\thefootnote}{$\star$}

\renewcommand{\PaperNumber}{074}

\FirstPageHeading

\ShortArticleName{Snyder Space-Time: K-Loop and Lie Triple System}

\ArticleName{Snyder Space-Time: K-Loop and Lie Triple System\footnote{This paper is a
contribution to the Special Issue ``Noncommutative Spaces and Fields''. The
full collection is available at
\href{http://www.emis.de/journals/SIGMA/noncommutative.html}{http://www.emis.de/journals/SIGMA/noncommutative.html}}}

\Author{Florian GIRELLI}

\AuthorNameForHeading{F. Girelli}

\Address{School of Physics, The University of Sydney, Sydney, New South Wales 2006, Australia}
\Email{\href{mailto:girelli@physics.usyd.edu.au}{girelli@physics.usyd.edu.au}}

\ArticleDates{Received April 29, 2010, in f\/inal form September 13, 2010;  Published online September 24, 2010}

\Abstract{Dif\/ferent deformations
of the Poincar\'e symmetries have been identif\/ied for  va\-rious  non-commutative spaces (e.g.~$\kappa$-Minkowski, $\sll(2,R)$, Moyal).
We present here the deformation of the Poincar\'e symmetries related to  Snyder space-time. The notions of smooth ``K-loop'', a non-associative generalization of Abelian Lie groups, and its inf\/initesimal counterpart given by the  Lie triple system are the key objects in the construction.}


\Classification{17C90;  81T75}

\renewcommand{\thefootnote}{\arabic{footnote}}
\setcounter{footnote}{0}

\section{Introduction}
\label{intro}

Snyder space-time has been introduced in 1947, and is one of the f\/irst examples of  non-com\-mu\-ta\-ti\-ve geometry \cite{snyder}. The physical features of a quantum f\/ield theory living on this space are not very well known since there have been very few attempts to construct such f\/ield theory on this space (see  \cite{elias} and references therein).  It is only recently that a star product and  a~scalar f\/ield theory have been constructed  \cite{girliv, girliv2}. Using a dif\/ferent route and doing the analysis at f\/irst order, \cite{meljanac}~reached similar results.  In both of these works, the scalar f\/ield action is shown to be invariant under a new type of deformed Poincar\'e symmetries. The main feature of this deformed symmetry is the non-(co-)associativity, which explains why this deformation did not appear when considering the (co-)associative deformations of the Poincar\'e group \cite{zak}.
In this perspective, the natural question to ask is  \textit{what is the algebraic structure or ``quantum group'' encoding the deformation of the Poincar\'e symmetries consistent with the Snyder non-commutativity}? We  present here the answer to this question by def\/ining a new type of quantum group.

Before explaining the strategy we shall use in the Snyder case, let us recall the construction in the  $\ka$-Minkowski non-commutative space which is well understood and can serve as a~gui\-ding line. In this case, the  relevant deformation of the Poincar\'e symmetries, the $\kappa$-Poincar\'e deformation~\mbox{\cite{lukierski, ruegg}},  is based on the Iwasawa decomposition of $\SO(p,1)\sim \AN_p\,.\,  \SO(p-1,1)\sim  \AN_p\,\double\, \SO(p-1,1)$ \cite{majid, iwasawa}. The (Abelian nilpotent) group $\AN_p$ is interpreted  as momentum space when dealing with a  scalar f\/ield theory living in $\ka$-Minkowski. The deformation of the translation symmetry can be traced back to the non-Abelian $\AN_p$ group structure. The $\ka$-Poincar\'e (or bicrossproduct) algebra is encoded by $\bb=\cc(\AN_p)\,\bicross\, k\SO(p-1,1)$, where the algebra of functions over $\AN_p$ coacts on the group algebra $k\SO(p-1,1)$, and $k\SO(p-1,1)$ acts on $\AN_p$. This quantum group encodes the deformation of the Poincar\'e Hopf algebra $\cc(\R^p)\,\rtimes\, k\SO(p-1,1)\stackrel{\ka}{\longrightarrow} \cc(\AN_p)\,\bicross\, k\SO(p-1,1)$, by implementing the Planck scale $\ka$. The $\ka$-Minkowski space is encoded into the coordinate operators $X_\mu$ which satisfy the same commutator relations as the Lie algebra~$\an_p$ of~$\AN_p$. Using the Weyl map, it is possible to construct a $*$-product between c-number coordinates~$x_\mu$ which  satisfy  the same commutation relations as~$\an_p$~\cite{laurent-kowalski}.

We can follow a similar route to construct the relevant non-commutative space associated to Snyder space-time.  The Snyder deformation will be based on the  decomposition  $ \SO(p-1,1)\sim L  \,.\, \SO(p-1,1)$, where  $L$ is not  a group but a \textit{K-loop}, a  non-associative generalization of Abelian groups \cite{kloop}. In this case, the deformation of the translation symmetry will be related to the~K-loop structure. The goal of the paper is to construct the quantum group $\sss= \cc(L)\rtimes k\SO(p-1,1)$~-- where the group algebra $k\SO(p-1,1)$ acts on the algebra of functions $\cc(L)$ on the K-loop~-- with the relevant Hopf structures. This quantum group  can  be seen once again as a deformation of the Poincar\'e Hopf algebra $\cc(\R^p)\,\rtimes\, k\SO(p-1,1)\stackrel{\ka}{\longrightarrow} \cc(L)\rtimes k\SO(p-1,1)$, but not in the  Hopf or quasi-Hopf algebra setting, since  the antipode of $\sss$ will not be antimultiplicative. This means in particular that this deformation has not been identif\/ied in Zakrzewski's classif\/ication of Poincar\'e deformations \cite{zak}. We will show that this quantum group $\sss$   encodes the deformed Poincar\'e symmetry identif\/ied in \cite{girliv, girliv2}. The inf\/initesimal structure of a K-loop is a Lie triple system and we will explain how   Snyder non-commutative structure is naturally encoded in such  structure. \textit{Lie triple systems provide therefore a new type of non-commutative spaces}.  This is the other main result of this paper.  As a side result,   one  recovers that the $*$-product def\/ined in~\mbox{\cite{girliv, girliv2}} is actually a realization of the Lie triple structure.

 In the f\/irst section we  recall  the dif\/ferent notions of loop which can be found in the standard literature, emphasising on the K-loop case. In the second section we introduce the main result of the paper that is the construction of the quantum group encoding the symmetries of Snyder spacetime. In the third section, we show how to construct a scalar f\/ield action invariant under such symmetry, using the method described in~\cite{noui}. In the fourth section, we  recall the inf\/inite\-si\-mal notion of a smooth K-loop given by Lie triple system and explain how this structure can be interpreted as encoding Snyder  non-commutative geometry. The concluding section sets a~list of interesting points  to explore.

\section{K-loop: a review}\label{sec:loop}

We recall in this section the basic def\/initions of the most studied loops with particular emphasis  on the notion of K-loop,   a (midely) non-associative generalization  of Abelian group.
 Most of the  content of this section (except Proposition~\ref{thm}, Def\/inition~\ref{lie group} and Example~\ref{example 2}) can be found in the standard reference on K-loops \cite{kloop} (see also \cite{sabinin:book} for further references on the notion of loops).

\begin{definition}[quasi-group and loop]\label{quasigroup}
A \textit{quasigroup} $(S, \cdot)$ is a set $S$ with a binary operation~``$\cdot$'' such that for each $a$ and $b$ in $S$, there exist unique elements $x$ and $y$ in $S$ such that $
a\cdot x = b$, $
   y\cdot a = b$.
 A quasigroup is a \textit{loop} $L$ if it has also an identity $e$ such that
$ e\cdot a=a\cdot e=a.$
\end{definition}
Note that this implies in particular that in a loop, we have a unique identity element and that the left and right inverse are unique.

\begin{definition}[left Bol loop]\label{bol loop}
Let $L$ be a loop.
The \textit{left Bol identity} is
\begin{gather*}
a\cdot(b\cdot(a\cdot c)) = (a\cdot(b\cdot a))\cdot c,  \qquad \forall\, a,b,c \in L.
\end{gather*}
The \textit{right Bol identity} is
\begin{gather*}
(c\cdot a)\cdot b)\cdot a = c\cdot((a\cdot b)\cdot a)),\qquad \forall \, a,b,c \in L.
\end{gather*}
A loop $L$ is said to be a \textit{left Bol loop} (resp.\ right Bol loop) if it satisf\/ies the left Bol identity (resp.\ the right Bol identity).
\end{definition}

\begin{proposition}[alternative loop]\label{alternative}
Let $L$ be a left Bol loop, then it is left alternative, that is
\begin{gather*}
a\cdot (a\cdot b)= a^2\cdot b= (a\cdot a)\cdot b.
\end{gather*}
\end{proposition}

This proposition follows from the left Bol property with $b=1$. The consistency relationships between the product and inverse map of a loop will characterize dif\/ferent types of loop.

\begin{definition}[inverse properties]\label{inverse}
Let $L$ be a loop.
\begin{itemize}\itemsep=0pt
\item The \textit{automorphic inverse property} (AIP)  is
$(a\cdot b)\mone = a\mone\cdot b\mone$, $\forall\, a,b \in L.$

\item The \textit{inverse property} (IP) is
$ (a\cdot b)\mone= b\mone\cdot a\mone$, $\forall\, a,b \in L.$

\item The  \textit{left inverse property} (LIP) is
$\exists\, a\mone\in L$, $ a\mone\cdot (a\cdot b) = b$, $\forall\, a,b \in L.$
\item The \textit{right inverse property} (RIP) is $ \exists \,b\mone\in L$, $(a\cdot b)\cdot b\mone = a$, $\forall\, a,b \in L.$
\end{itemize}
\end{definition}
Note that the LIP together with the RIP gives the IP.
Let us give now some examples of loops.
\begin{definition}[Lie group]\label{lie group}
A  \textit{group}  is  a  loop satisfying the \textit{inverse  property} and
\begin{gather*}
a\cdot (b \cdot c)=( a\cdot b) \cdot c= a\cdot b \cdot c.
\end{gather*}
\end{definition}
A group is therefore in particular  a left and right Bol loop.
\begin{definition}[Moufang loop]\label{moufang loop}
A  \textit{Moufang loop } is  a left and right Bol loop. It satisf\/ies in particular the \textit{inverse property}.
\end{definition}

\begin{definition}[left K-loop]\label{k loop 0}
A  \textit{left    K- loop} (or left Bruck loop)  is  a left Bol loop satisfying the \textit{automorphic inverse  property}.
\end{definition}

In \cite{hopfloop}  the authors focused on the Hopf  structures related to the Moufang case, which are a~natural non-associative generalization of non-Abelian groups. We are interested instead in the K-loop case.
In the following, we shall omit the term ``left'' since we shall always consider left  K-loops, unless specif\/ied otherwise. To generate loops, one can consider the factorization of Lie groups.

\begin{proposition}\label{thm}
Consider the group $\SO(p,1)$. The  decomposition $\SO(p,1)\sim L \,.\, \SO(p-1,1)$ provides a unique decomposition
\begin{gather*}
g=ah, \qquad g\in \SO(p,1), \qquad h\in \SO(p-1,1), \qquad a\in L.
\end{gather*}
The product of $\SO(p,1)$ induces a product ``$ \cdot $'' in $L$
\begin{gather*}
ab= (a\cdot b)h_{ab} , \qquad \forall\,  a,b\in L, \quad h_{ab}\in\SO(p-1,1).
\end{gather*}
The  groupoid $L$ is a smooth K-loop with product ``$\cdot$''. Its inverse map coincides with the group inverse map.
\end{proposition}

The same proposition holds for the Cartan decomposition $\SO(p,1)\sim L'\,.\, \SO(p)$,  and the decomposition $\SO(p)\sim L''\,.\, \SO(p-1)$, i.e.\ both $L'$ and $L''$ are K-loops. The proof for these cases is given in \cite{kloop}. This proposition and its variants for the other decompositions can also be proved using the notion of   Lie triple systems \cite{kikawa}.

There is a weak form of commutativity and associativity present in a K-loop. When constructed from a group factorization as in the above decompositions, it takes the shape
 \begin{gather}
 a\cdot b= h_{ab}  (b\cdot a)  h_{ab}\mone, \label{gyrocom}\\
 a\cdot (b\cdot c)= (a\cdot b)\cdot\big(h_{ab}  c   h_{ab}\mone\big).\nn 
\end{gather}
Hence K-loops can be seen as non-associative generalization of Abelian groups.  To conclude this section, let us give two examples of K-loops.

\begin{example}\label{example 1}
Consider the orthochronous Lorentz group $\SO^+(3,1)$ and its Cartan decomposition $\SO^+(3,1)\sim L'\,.\, \SO(3)$.  $L'$  is  a K-loop \cite{ungar, sabinin} which can be identif\/ied to the space of 3D speeds in Special Relativity, and the loop product encodes the speed addition used when changing reference frame. Let us recall its structure. $L'$ is isomorphic to the 3d upper hyperboloid $H^+_3$. This space can be embedded in $R^4$ as   $H^+_3=\lbrace V_\mu\in\R^4, \  V^\mu V_\mu = 1, \  V_0>0 \rbrace$ and is generated by the  boost  $a= e^{i\frac{\eta}{2} \vec b\cdot \vec K}$, where $K_i\in \so(3,1)$ are the boost generators, $\eta$ is the boost angle,  $\vec b$~is the boost vector ($\vec b\,{}^2=1$), and we are using the spinorial representation of $\SL(2,\C)$ \cite{SR, girliv2}. The product of boosts gives the loop product:
\begin{gather}
  e^{i\frac{\eta_1}{2} \vec b_1\cdot \vec K}e^{i\frac{\eta_2}{2} \vec b_2\cdot \vec K} =  e^{i\frac{\eta_{12}}{2} \vec b_{12}\cdot \vec K} R_{\eta_i,\vec b_i}\Longleftrightarrow a_1a_2= a_{12}h_{12}  \nn\\
\hspace*{30mm} \downarrow  \label{boost product}\\
  e^{i\frac{\eta_1}{2} \vec b_1\cdot \vec K}\cdot e^{i\frac{\eta_2}{2} \vec b_2\cdot \vec K} \equiv  e^{i\frac{\eta_{12}}{2} \vec b_{12}\cdot \vec K}\Longleftrightarrow a_1\cdot a_2= a_{12}.    \nn
\end{gather}
The rotation $R_{\eta_i,\vec b_i}$ is  called the Thomas precession. The 3d speed $\vec v$ in Special Relativity is given in terms of the coordinates  on $H_3^+$ by
\begin{gather*}
\vec v(a)\equiv \vec v= c \tanh\eta  \vec b= c \frac{\vec V}{V_0} ,
\end{gather*}
where $c$ is the speed of light. The loop product \eqref{boost product} induces the sum of speeds:
\begin{gather*}
\vec v(a_1)\oplus \vec v(a_2)\equiv\vec v_1\oplus \vec v_2 =  c\tanh \eta_{12}   \vec b_{12}=\vec v(a_1\cdot a_2)\nn\\
 \phantom{\vec v(a_1)\oplus \vec v(a_2)}{} =  \frac{1}{1+\frac{\vec v_1\cdot \vec v_2}{c^2}}\left(\left(1+ \frac{\gamma_1}{1+\gamma_1}\frac{\vec v_1\cdot \vec v_2}{c^2}\right)\vec v_1 +  \frac{1}{\gamma_1}\vec v_2\right), \qquad \gamma_1= \frac{1}{\sqrt{1-\frac{\vec v^2_1}{c^2}}}.\nn
\end{gather*}
It is not dif\/f\/icult to check that the inverse $\ominus$ coincides with the usual inverse $\ominus\vec v= -\vec v$ and  this addition satisf\/ies the automorphic inverse property $-(\vec v_1\oplus \vec v_2)= (-\vec v_1)\oplus (-\vec v_2)$.  This K-loop  can also be interpreted as momentum space in which case one reconstructs an Euclidian Snyder space-time \cite{girliv, girliv2}.
\end{example}

\begin{example}\label{example 2}
Consider the  decomposition $\SO(4,1)\sim L\,.\, \SO(3,1)$. In this case $L$ is isomorphic to the de Sitter space $dS$, which can be embedded in $\R^5$ as $dS=\lbrace \pi_A\in\R^5, \, \pi^A \pi_A = -1 \rbrace$. It is generated by the de Sitter boosts $a=e^{i\frac{\eta}{2} B\cdot J_4 }$, where $J_4\equiv J_{4\mu}\in \so(4,1)$ ($\mu=0,\dots,3$), $\eta$~and~$B^\mu$ are  the angle and the boost vector ($B^2=-1$) respectively and we are using the spinorial representation of $\SO(4,1)$. The de Sitter space is covered by two coordinate charts given by $\pi_4>0 $ and $\pi_4<0$.  As in the previous example, we restrict to the upper part of the de Sitter space, i.e.\ the sector with $\pi_4>0$ (which is stable under the Lorentz action).  Snyder used the de Sitter space to def\/ine momentum space and  momentum is given in terms of the coordinates on de Sitter space \cite{snyder}
\begin{gather}\label{snyder coord}
p_\mu(a)\equiv p_\mu= \ka \tanh\eta  B_\mu=\ka \frac{\pi_\mu}{\pi_4},
\end{gather}
$\ka$ is the Planck mass ($\hbar=c=1$ in this example). He did not def\/ine the notion of momenta addition, but we can use the K-loop structure to def\/ine it
\begin{gather*}
e^{i\frac{\eta_1}{2}  B_1\cdot  J_4}e^{i\frac{\eta_2}{2}  B_2\cdot J_4} =  e^{i\frac{\eta_{12}}{2} B_{12}\cdot } \Lambda_{\eta_i,\vec B_i} \Longleftrightarrow a_1a_2= a_{12}h_{12}   \nn\\
\hspace*{33mm} \downarrow \\ 
e^{i\frac{\eta_1}{2}  B_1\cdot J_4}\cdot e^{i\frac{\eta_2}{2}  B_2\cdot J_4} \equiv   e^{i\frac{\eta_{12}}{2} B_{12}\cdot J_4} \Longleftrightarrow a_1\cdot a_2= a_{12},  \nn
\end{gather*}
where $\Lambda_{\eta_i,\vec B_i}$ is Lorentz transformation determined by the coordinates $(\eta_i,B_i)$. Using the expression \eqref{snyder coord}, we obtain the addition of momenta $ p_\mu(a_1)\oplus p_\mu(a_2)\equiv(p\oplus q)_\mu= \ka\tanh \eta_{12}   \vec B_{12}=p_\mu(a_1\cdot a_2)$  given in the appendix. Since it is quite complicated we give here the f\/irst order contribution in $\ka^2$ (with $p\cdot q=p^\mu q_\mu$, $p^2=p^\mu p_\mu$)
\begin{gather}\label{snyder sum1}
(p\oplus q)_0 = p_0+q_0+\frac{1}{2\ka^2} \left((p_0+2q_0)  p\cdot  q  +    q_0   p^2\right)+O\left(\frac{1}{\ka^3}\right),\\
(p\oplus q)_i = p_i+q_i+\frac{1}{2\ka^2 }\left((p_i+2q_i)  p\cdot  q   +(q_i-p_i) p_0q_0  -  p_i  q_0^2+  q_i   \vec p\,{}^2\right)+O\left(\frac{1}{\ka^3}\right). \label{snyder sum2}
\end{gather}
Once again, it is easy to check that $-p$ is the inverse of the addition $\oplus$. A direct calculation shows that the addition $\oplus$ satisf\/ies the automorphic inverse property law, that is $-(p_1\oplus p_2)= (-p_1)\oplus (-p_2)$.
\end{example}

\section{Snyder quantum group}
In this section we present the main result of this paper, that is the construction of the quantum group encoding the symmetries of Snyder spacetime. Firstly, we  def\/ine the notion of \textit{K-Hopf loop}, which  will be used to def\/ine the K-loop algebra $kL$, the analogue of the group algebra. Secondly, we def\/ine K-Hopf coloop which is the dual notion of K-Hopf loop. A specif\/ic example will be the algebra of functions on the K-loop $\cc(L)$. Finally, we introduce the Snyder quantum group $\cc(L)\rtimes k\SO(p-1,1)$  built out from cross product of the Lorentz group algebra with the algebra of functions on the K-loop.

\subsection{K-Hopf loop}
We def\/ine here the relevant  Hopf structure for the  K-loop. We use the Sweedler notation for the coproduct $\cop a= a_{(1)}\ot a_{(2)}$.
\begin{definition}[K-Hopf loop]\label{k loop} Let $k$ be a f\/ield of characteristic $\neq 2,3$ and $a,b\in \hh$.
A \textit{K-Hopf loop} is a unital algebra $\hh$, equipped with algebra homomorphisms $\cop:\hh\dr \hh\ot \hh$, $\epsilon:\hh\dr k $ forming a coassociative coalgebra, and a map $S:\hh\dr \hh$ such that
\begin{gather}
  a_{(1)}\cdot(b\cdot (a_{(2)}\cdot c))= (a_{(1)}\cdot (b\cdot a_{(2)}))\cdot c, \label{hopf bol}\\ 
  a_{(1)}\cdot (S( a_{(2)})\cdot b)= b= S(a_{(1)})\cdot (a_{(2)}\cdot b). \label{left inverse1}
\end{gather}
The  equations   \eqref{left inverse1} are the alter-ego of the left inverse property, whereas \eqref{hopf bol} is related to the left Bol property.
The antipode $S$ is furthermore  ``multiplicative'' and ``comultiplicative''
\begin{gather*}
  S(a\cdot b)= S(a)\cdot S(b),\\
  \cop (S(a))= S(a_{(1)})\ot S(a_{(2)}).
\end{gather*}
\end{definition}
This antipode is  very dif\/ferent to the one met in a Hopf algebra  \cite{majid} and in  a  Moufang Hopf quasigroup \cite{hopfloop}, since in these cases it is  ``antimultiplicative'' and ``anticomultiplicative''.
\begin{proposition}\label{propriete bol hopf loop}
A  K-Hopf loop satisfies
\begin{gather}
  a_{(1)}\cdot (a_{(2)}\cdot b)= (a_{(1)}\cdot a_{(2)})\cdot b, \label{hopf alternative}\\
  S(a_{(1)})\cdot a_{(2)}= a_{(1)} \cdot S(a_{(2)}).\label{right inverse}
\end{gather}
\end{proposition}
\begin{proof}
\eqref{hopf alternative} says that the K-Hopf loop is left alternative \cite{hopfloop}. This is obtained from \eqref{hopf bol} by setting $b=1$.
\eqref{right inverse} is obtained from \eqref{left inverse1}  applied on $a\ot 1$.
\end{proof}

The group  algebra $kG$ for a group $G$ provides a natural example of Hopf algebra \cite{majid}. In a~similar way, the K-loop algebra $kL$ provides  an  example of  K-Hopf loop.
\begin{proposition}
If $L$ is a K-loop, then $\hh=kL$ is a K-Hopf loop with linear extension of the product and on the basis elements $a,b$ we have
\begin{gather}
  m(a\ot b)= a\cdot b, \qquad  \epsilon (a)= 1 , \qquad \cop a= a\ot a, \qquad  S(a)= a\mone,
\end{gather}
and the unit is given by $e$.
\end{proposition}
The proof is straightforward. We notice in particular that $kL$ is both co-commutative and co-associative.  The inf\/initesimal version of smooth K-loops  is a Lie triple system~\cite{kikawa}. The enveloping algebra for a Lie triple system has been constructed in~\cite{spanish}. This is another example of  K-Hopf loop and can be seen as the inf\/initesimal version of $kL$, when $L$ is a smooth K-loop.

\subsection{K-Hopf coloop}

Once we have linearised the concept of K-loop into the K-loop algebra $kL$, one can reverse the arrows on all the maps, and obtain a K-Hopf coloop. We note $m$ the multiplication.
\begin{definition}[K-Hopf coloop]\label{coK hopf} Let $k$ be a f\/ield of characteristic $\neq 2,3$ and $f,f_i\in \hh$.
A~\textit{\mbox{K-Hopf} coloop} is a unital associative  algebra $\hh$, equipped with counital algebra homomorphisms $\cop:\hh\dr \hh\ot \hh$, $\epsilon:\hh\dr k $  and a linear map $\iota:\hh\dr \hh$
\begin{gather*}
  m(f_{(21)}\ot f_{(1)})\ot f_{(221)}\ot f_{(222)}=m(f_{(121)}\ot f_{(11)})\ot f_{(122)}\ot f_{(2)}, \\ 
 (m\ot \i)(\iota\ot \i \ot \i)(\i \ot \cop)\cop= 1\ot \i 
 = (m\ot \i)(\i\ot \iota \ot \i)(\i \ot \cop)\cop ,
\\
 \iota(m(f_1\ot f_2))= m(\iota f_1\ot \iota f_2), \\
 \cop (\iota f)= \iota f_{(1)}\ot \iota f_{(2)}.
\end{gather*}
By counital, we mean $ (\i\ot\epsilon)\cop=(\epsilon\ot \i)\cop=\i$.
\end{definition}
A K-Hopf coloop satisf\/ies the analogue of Proposition \ref{propriete bol hopf loop}, where the arrows are reversed.
\begin{proposition}\label{propriete bol co hopf loop}
A  K-Hopf coloop satisfies
\begin{gather*}
  f_{(1)}\cdot f_{(21)}\ot f_{(22)}= f_{(11)}\cdot f_{(12)}\ot f_{2}, 
  \\   
  m(\iota\ot \i)\cop= \id \epsilon = m (\i\ot \iota)  \cop . 
\end{gather*}
\end{proposition}

\begin{proof}
From \eqref{hopf alternative}, we have for $f\in k(L)$, $a,b\in L$,
\begin{gather*}
(f,a_{(1)}\cdot (a_{(2)}\cdot b))=(f, (a_{(1)}\cdot a_{(2)})\cdot b)\nonumber\\
 \qquad{} \Leftrightarrow\   (f_{(1)}\cdot f_{(21)}\ot f_{(22)}, a\ot b)= (f_{(11)}\cdot f_{(12)}\ot f_{2} , a\ot b).
\end{gather*}
On the other hand, we have from \eqref{right inverse}
\begin{gather*}
(f,a_{(1)}Sa_{(2)})=  (f,Sa_{(1)}a_{(2)}) = \epsilon(f)\epsilon(a)\Leftrightarrow ( m(\iota\ot \i)\cop f,a)= ( m (\i\ot \iota)  \cop f,a) .\tag*{\qed}
\end{gather*}
\renewcommand{\qed}{}
\end{proof}

As an   example, we  construct  the K-Hopf coloop $k(L)$ -- which is identif\/ied to the algebra of functions on $L$ -- by duality, from the K-Hopf loop $kL$. We use the duality $(f,a)\equiv f(a)$, $\forall f\in k(L)$, $a$ basis element in $kL$, and it  is  extended by linearity to all elements in $kL$.
\begin{proposition}
If $L$ is a K-loop, then $\hh=k(L)$ is a K-Hopf coloop, with
\begin{gather}
(m(f_1\ot f_2),a)=f_1(a)f_2(a), \qquad \epsilon (f)=f( e), \nonumber\\
(\cop f, a\ot b)= f(a\cdot b), \qquad (\iota f,a)=f( a\mone), \label{kcoloop}
\end{gather}
with $a$, $b$ elements in the basis of $kL$ and $f_i\in k(L)$, and the unit function is $\id$ such that $\id(a)=1$, $\forall\, a\in L$.
\end{proposition}

The proof is straightforward. We notice that $k(L)$ is a commutative and associative algebra, but is not co-commutative nor co-associative.

The general Hopf coloop satisf\/ies similar properties as in Proposition \ref{propriete bol hopf loop}, which can be simply shown by reversing the arrows of the  maps.
\begin{example}\label{example 2 bis}
Consider the K-Hopf loop generated from $L\sim \SO(4,1)/\SO(3,1)$   as given in Example \ref{example 2}. The dual structure is given by the set of functions $\cc(L)$. The coordinate functions $p_\mu\in \cc(L)$ is a natural example to consider. The coproduct $\cop p_\mu$ is then the dual of  the sum~\eqref{snyder sum1},~\eqref{snyder sum2}
\begin{gather*}
\langle (\cop p_\mu) ;a_1 \ot a_2\rangle= \langle p_{\mu}; a_1 \cdot  a_2\rangle= p_\mu(a_1 \cdot  a_2) = (p\oplus q)_\mu, \qquad a_i\in L.
\end{gather*}
\end{example}

\subsection{Snyder Hopf loop}

We are now presenting the algebraic structure that is encoding  the  deformation of the Poincar\'e symmetry  consistent with the Snyder space. We focus on the f\/ield $k=\C$.  Considering the  decomposition $\SO(p,1)\sim L \,.\,\SO(p-1,1) $, $L$ is isomorphic to the de Sitter space. There is a~natural action of $\SO(p-1,1)$ on $L$ given in terms of the adjoint action $a\dr u a  u\mone=u\act a$, $\forall \, a\in L$,  $u\in \SO(p-1,1)$. This action is naturally lifted to the action of $k\SO(p-1,1)$ on $k(L)$,  with $(u\act f,a)\equiv f(u\act a)$ and $u$ basis element of $k\SO(p-1,1)$. We recall that the group Hopf algebra $k\SO(p-1,1)$ is specif\/ied by the coproduct, antipode, counity and unity respectively def\/ined on basis element $u\in\SO(p-1,1)$ (and extended by linearity),
 \begin{gather}\label{group algebra}
{\bf \cop} u=u\ot u, \qquad {\bf S}u=u\mone, \qquad {\bf \epsilon}(u)=1, \qquad e.
 \end{gather}

\begin{proposition}\label{module}
$k(L)$ is a $k\SO(p-1,1)$-module algebra and coalgebra,  i.e.
\begin{gather}
 u\act(f_1. f_2)= m(\cop u\act (f_1\ot f_2)), \qquad  u\act \id= \epsilon(u)\id= \id, \label{first}\\
  \cop(u\act f)= {\bf\cop} u \act \cop f,  \qquad \epsilon(u\act f)={\bf\epsilon}(u)\epsilon(f)=\epsilon(f) , \label{second}
\end{gather}
with $u$ basis elements of $k\SO(p-1,1)$ and we used \eqref{group algebra}. This is naturally extended by linearity. We have moreover
\begin{gather}
\iota(u\act f)= u\act (\iota f).\label{third}
\end{gather}
\end{proposition}

\begin{proof}
We shall consider $a$, $b$ basis elements of $kL$ before extending naturally by linearity. First we prove~\eqref{first},
\begin{gather*}
 (u\act (f_1.f_2),a) =  ((f_1.f_2),u\act a)= (f_1\ot f_2,\cop (u\act a) ) = ((f_1\ot f_2, (u\act a) \ot (u\act a) )) \\
\phantom{(u\act (f_1.f_2),a)}{}
 =   ((f_1\ot f_2, (\cop u)\act (\cop a)) ))=  (\cop u\act (f_1\ot f_2),\cop a)\nn\\
\phantom{(u\act (f_1.f_2),a)}{} =  (m(\cop u\act (f_1\ot f_2)),a).
\end{gather*}
We have moreover that
\[
 ( u\act \id, a) =(\id, u\act a)= \epsilon(u\act a)= 1= \epsilon(a)=(\id,  a),
\]
where $\epsilon(u\act a)$ is the counity on $kL$.
Then we prove \eqref{second}
\begin{gather*}
 (\cop (u\act f),a\ot b) =  (u\act f,a\cdot b)= (f,u\act(a\cdot b)) = (f,(u\act a)\cdot (u\act b))\nn\\
\phantom{(\cop (u\act f),a\ot b)}{}
= (\cop f,(u\act a)\ot (u\act b) ) = ( (\cop u)\act (\cop f), a\ot b).
\end{gather*}
We have also
\[
\epsilon(u\act f)= (u\act f, e)= (f,e    u \textbf{S} u  )= (f,e)\epsilon(u)= \epsilon(f).
\]
Finally the  proof of  \eqref{third} goes as follows
\[
(\iota(u\act f), a) =  \big(u\act f,a\mone\big)= \big(f,u\act a\mone\big)= (\iota f, u\act a)=(u\act \iota f,a),
\]
where we used that the inverse of $L$ and $\SO(p-1,1)$ actually coincides with the inverse in~$\SO(p,1)$.
\end{proof}

Since $k(L)$ is  a $k\SO(p-1,1)$ module and co-module, we can consider the semi-direct pro\-duct~\cite{majid}.

\begin{definition}[Snyder Hopf loop]\label{snyder loop}
The \textit{Snyder  Hopf loop} $\sss\equiv \cc(L)\rtimes k\SO(p-1,1)$ is given
 in terms of the vector space $\cc(L)\otimes k\SO(p-1,1)$, and respectively, product, coproduct, unit, counit, antipode and complex conjugate
\begin{alignat*}{3}
& \textrm{Product} \quad && (f_1\ot u_1)(f_2\ot u_2)= (f_1. ({u_1}\act f_2)\ot u_1u_2), & \\
& \textrm{Coproduct} \quad && \cop(f\ot u)= (f_{(1)}\ot u)\ot (f_{(2)}\ot u), & \\
& \textrm{Unit and counit} \quad && \id \ot e , \quad \epsilon(f\ot u)= f(e),& \\
& \textrm{Antiopde}&&\ss(f\ot u)= \big({u\mone}\act \iota  f\ot u\mone\big),& \\
& \textrm{Complex structure} \quad &&(f\ot u)^*= (\overline{u\mone\act f}\ot u\mone).&
\end{alignat*}
 These relations are naturally extended by linearity when $u$ is a general element of $k\SO(p-1,1)$.  The coproduct  $\cop f = f_{(1)}\ot f_{(2)}$ is  given in~\eqref{kcoloop}.
\end{definition}

This construction can be extended to the other cases generated by the  decompositions we mentioned above: $\cc(L'')\rtimes k\SO(p-1)$ and $\cc(L')\rtimes k\SO(p) $ are also Snyder Hopf loops.
We have the following proposition
\begin{proposition}\label{snyder}
The antipode in the Snyder Hopf loop is antimultitplicative but comultiplicative
\begin{gather}
  \ss\left((f_1\ot u_1)(f_2\ot u_2))\right)= \ss(f_2\ot u_2)\ss(f_1\ot u_1),\nonumber \\
  \cop \ss = (\ss\ot \ss )\cop,\nonumber \\
  m(\ss\ot \i)\cop=m(\i\ot \ss)\cop=\id \epsilon\ot e \epsilon .\label{last3}
\end{gather}
\end{proposition}

\begin{proof}
\begin{gather}
 \ss\left((f_1\ot u_1)(f_2\ot u_2))\right) =  \ss(f_1 .(u_1\act f_2)\ot u_1u_2)=
 \big((u_1u_2)\mone\act \iota(f_1 .(u_1\act f_2))\ot  (u_1u_2)\mone\big)\nn\\
 \phantom{\ss\left((f_1\ot u_1)(f_2\ot u_2))\right)}{}
 =  \big((u_1u_2)\mone\act (\iota f_1 .(u_1\act \iota f_2))\ot  (u_1u_2)\mone\big)\nn\\
\phantom{\ss\left((f_1\ot u_1)(f_2\ot u_2))\right)}{}
   =   \big((u_1u_2)\mone\act (\iota f_1) .(u_2\mone\act (\iota f_2))\ot  (u_1u_2)\mone\big),\label{previous}
\end{gather}
where we used Proposition~\ref{module}, and the properties of the action
\begin{gather*}
 \ss(f_2\ot u_2)\ss(f_1\ot u_1) =  \big(u_2\mone \act (\iota f_2)\ot  u_2\mone\big)\big( u_1\mone \act (\iota f_1)\ot  u_1\mone\big) \nn\\
 \phantom{\ss(f_2\ot u_2)\ss(f_1\ot u_1)}{}
  =  \big(u_2\mone \act (\iota f_2). (u_2\mone u_1\mone) \act (\iota f_1)\ot  u_2\mone u_1\mone\big).
\end{gather*}
This is equal to  \eqref{previous}  if we recall that the algebra $k(L)$ is commutative. Finally we show that the the antipode is comultiplicative
\begin{gather*}
 \cop \ss (f\ot u)  = \cop \big(u\mone\act \iota f\ot  u\mone\big)= \big(\big(\cop u\mone \big)\act (\cop \iota f)\ot  \cop u\big)
 \\
 \phantom{\cop \ss (f\ot u)}{}
 = \big(\big(\cop u\mone \big)\act( (\iota\ot \iota )\cop  f)\ot  \cop u\big)
 =   (\ss\ot \ss )\cop(f\ot u),
\end{gather*}
where we used the consistency relations between the coproduct and antipode for $k(L)$ and $k\SO(p,1)$ inherited from Proposition \ref{module}.
The proof of \eqref{last3} follows directly from Proposi\-tion~\ref{propriete bol co hopf loop}.
\end{proof}

\section{Scalar f\/ield theory}\label{sec:field theory}

Once we have def\/ined the fundamental algebraic structure, we can proceed to construct a scalar f\/ield theory, invariant under such symmetry group, following the same route as  \cite{noui}. We focus here on the 4d Lorentzian case, i.e.\ $L\sim dS$ and we are going to show how to recover the results of \cite{girliv, girliv2}.

As in Example~\ref{example 2}, we focus on the upper part of the de Sitter space $L^+$. We consider therefore the Snyder Hopf loop $\sss^+=\cc(L^+)\rtimes \, k\SO(3,1)$ which is naturally obtained from $\sss$ by restricting the set of functions $\cc(L)$ to $\cc(L^+)$. This restriction is consistent with the Hopf structures given in Def\/inition~\ref{snyder loop}.

The scalar f\/ield is an element of the dual $\dd^*$ of  the coset $k(L^+)\sim \sss^+/k\SO(3,1)$, which can be specif\/ied as the set of distributions with compact support $k^*(L^+)$. $\forall\, f\ot u \in \sss$, $v\in \SO(3,1)$
\begin{gather*}
\dd^*=  \lbrace F\in {\sss^+}^* \left|
\langle(f\ot u)(\id\ot v),F \rangle= \langle f\ot u, F\rangle\right.\rbrace \\
\qquad{} \sim  k^*(L^+)=\lbrace \Phi| \langle f,\Phi\rangle= \langle f\ot e,F\rangle, \, f\in k(L^+)\rbrace.
\end{gather*}
This set of distributions is equipped with the convolution product which makes $k^*(L^+)$ an algebra. The convolution product is constructed using the coproduct on $\sss^+$,
\[
\langle f,\Phi_1\circ \Phi_2\rangle\equiv \langle \cop_\sss( f\ot e), F_1\ot F_2\rangle= \langle \cop_{k(L)}f, \Phi_1\ot \Phi_2\rangle.
\]
The scalar f\/ield can be seen as a function on $L^+$, which is then interpreted as momentum space. Hence, it is  convenient to consider the subalgebra of distributions given by the algebra of functions (with compact support) $k(L^+)\subset \dd^*$. For $\phi_i\in k(L^+)$, we have
\[
\phi_1\circ\phi_2 (a)= \int [da_i]^2\phi_1(a_1)\phi_2(a_2)\delta\big(a\mone \cdot (a_1\cdot a_2)\big),
\]
where we are using  the measure $[da]=d^5\pi\,\delta (\pi_A\pi^A+1)\theta(\pi_4) $ on $L^+$, induced from the group decomposition and the loop product.    This convolution product between functions is used to generate the relevant terms in the scalar f\/ield action. The propagating term and the $\phi^3$-like interaction term are given as
\begin{gather}
\label{propa}
\Psi\circ \phi(e) = \int [da]^2 \big(p^2(a_1)+m^2\big)\phi(a_1)\phi(a_2)\delta(a_1\cdot a_2), \qquad \Psi(a)= \big(p^2(a)+m^2\big)\phi(a),\\
\label{interaction}
\phi\circ (\phi\circ\phi)(e) =  \int [da]^3\phi_1(a_1)\phi_2(a_2)\phi_2(a_2)\delta(a_1 \cdot (a_2\cdot a_3)),
\end{gather}
where the delta function encodes the generalization of the momentum conservation law and~$p_\mu(a)$ is the Snyder coordinates associated to the loop element $a$. Note that the convolution product is not  associative, since it is based on the loop product. It is therefore important to   keep track of the order of  the brackets. The action $\Sigma(\phi)$ for the scalar f\/ield $\phi$ reads explicitly
\begin{gather*}
\Sigma(\phi) = \Psi\circ \phi(e)+ \frac{\lambda}{3!} \phi\circ(\phi\circ\phi) (e)
= \int [dp]^2\, \phi(p_1)\big(p_1^2+m^2\big)\phi(p_2) \delta(p_1\oplus p_2)\\
\phantom{\Sigma(\phi)=}{}
+ \frac{\lambda}{3!} \int [dp]^3\,\phi(p_1)\phi (p_2)\phi(p_3) \delta(p_1\oplus(p_2\oplus p_3)).
\end{gather*}
In the second line, we have expressed the action in terms of the Snyder coordinates, in particular the measures reads $\ka^4[da]=[dp]=\demi d^4p\big(\big(1-\frac{p^2}{\ka^2}\big)\big)^{-\frac{5}{2}}$. The sum of momenta is given by~\eqref{snyder sum1},~\eqref{snyder sum2}.

We need to check that the Snyder Hopf loop $\sss^+$ encodes the right symmetry for this scalar action. We need therefore to def\/ine how the f\/ield $\phi$ is transforming under the quantum group action.  The symmetry action of $\sss^+$ on $k^*(L^+)$ is induced by its action on~$\sss^*$
\[
\langle f, (a\ot u)\act \phi\rangle \equiv  \langle f\ot e, (a\ot u)\act F \rangle
 =  \langle(a\ot u)^* ( f\ot e), F \rangle.
\]
This means concretely that the Lorentz group $\SO(3,1)$ sector acts by the adjoint action, whereas the $k(L^+)$ sector acts by multiplication. This last sector encodes the deformed translation symmetry
\begin{gather}\label{sym action}
\phi(a)\dr \phi(u\act a),\qquad  \phi(a)\dr f(a)\phi(a).
\end{gather}
To determine the exact realization of the translation symmetry, one needs to introduce space-time, through a Fourier transform. We consider therefore the c-numbers $x_\mu\in \R^4$ and the plane-wave $e^{ip\cdot x}$. To keep track of the modif\/ied sum of momenta
\eqref{snyder sum1}, \eqref{snyder sum2}, we introduce the modif\/ied product $*$ between plane-waves
\[
e^{ip_1\cdot x}*e^{ip_2\cdot x}\equiv e^{i(p_1\oplus p_2)\cdot x}.
\]
We can then introduce the Fourier transform of the f\/ield $\phi\in \cc(L^+)$
\[
\hphi(x)\equiv \int [da]\, e^{ip(a)\cdot x}\phi(a)= \int [dp]\, e^{ip\cdot x}\phi(p).
\]
The product between f\/ields is given by the $*$-product, which is the dual of the convolution product
\begin{gather*}
(\hphi_1*\hphi_2)(x) =   \int [dp]^2\, e^{ip_1\cdot x} *e^{ip_2\cdot x}\phi_1 (p_1)\phi_2(p_2)\\
\phantom{(\hphi_1*\hphi_2)(x)}{}
= \int [dp]^3\, e^{ip\cdot x} \phi_1 (p_1)\phi_2(p_2)\delta(-p\oplus (p_1\oplus p_2))
 = \int [dp]\, e^{ip\cdot x} \phi_1 \circ \phi_2(p).
\end{gather*}
With this Fourier transform, the scalar f\/ield action becomes
\[
\Sigma(\phi)= \int [dx]\left(- (\partial^\mu\hphi )* (\partial_\mu\hphi ) + m^2 \hphi*\hphi +\frac{\lambda}{3!}\hphi*(\hphi*\hphi)\right).
\]
The translation action is now given as $x\,\dr\, x+\varepsilon$ and the f\/ield $\phi(p)$ is transforming therefore as $\phi(p)\,\dr\, e^{ip\cdot \varepsilon}\phi(p)$, which gives us therefore $f(p)=e^{ip\cdot \varepsilon}$ for the symmetry action \eqref{sym action}.

To prove invariance of the action under $\sss^+$,
we also need to know how it is acting on convolution product. This is constructed naturally using the coproduct of $\sss^+$ and the coproduct on $k(L^+)$ as  in \eqref{kcoloop}
\[
(f\ot u)\act( \phi_1\circ \phi_2) \equiv ((f_{(1)}\ot u)\act \phi_1) \circ  ((f_{(1)}\ot u)\act \phi_2).
\]
It is straightforward to check that the action $\Sigma(\phi) $ constructed from the terms \eqref{propa} and \eqref{interaction} is then invariant under such transformations.  The coproduct of $\sss^+$ also provides    the  action of the Lorentz  group and the translations on a tensor product of f\/ields
\begin{gather*}
\phi(p_1)\phi(p_2) \,\dr\,  \phi(u\act p_1)\phi(u\act p_2),\qquad
\phi(p_1)\phi(p_2) \,\dr\,  e^{i(p_1\oplus p_2)\cdot \varepsilon} \phi(p_1)\phi(p_2).
\end{gather*}
The deformation of the translation symmetry is therefore naturally encoded in the loop product.

Before closing this section, we can evaluate explicitely the star product between dif\/ferent typical functions, using the properties of the Fourier transform we have introduced. For example, we can calculate the commutator of coordinates with the $*$-product  \cite{girliv, girliv2}
\begin{gather}\label{commutator zero}
\com{x_\mu,x_\nu}_*=0.
\end{gather}
The commutator of monomes of the f\/irst degree is zero. This is quite dif\/ferent than the usual non-commutative spaces. However, this does not mean that we have a commutative space. Indeed,  for example the commutator of a monome of the second degree with one of the f\/irst degree is not zero due to the non-associativity. As an example, let us look at the more complicated combination:
\begin{gather}\label{star triple product 1}
\langle x_\mu,x_\nu,x_\alpha\rangle \equiv  x_\mu*(x_\nu*x_\alpha) - x_\nu*(x_\mu*x_\alpha)
 =  -\frac{1}{\ka^2}(\eta_{\nu\alpha}x_\mu-  \eta_{\mu\alpha}x_\nu)=\frac{ i}{\ka^2}J_\mn\act x_\alpha,
\end{gather}
where $J_\mn\in\so(3,1)$ acts in the usual way on the coordinates~$x_\alpha$. By introducing a Weyl map, we are going to show in the next section  how this star product is the realization of the non-commutative structure implemented by a Lie triple system, the inf\/initesimal version of a smooth K-loop.

\section{Snyder space as a Lie triple system}

In the f\/irst subsection, we f\/irst recall the dif\/ferent examples of Bol algebras which are the inf\/initesimal version of the dif\/ferent smooth Bol loops we considered in Section \ref{sec:loop}. We focus on the Lie triple system case and show how given a Lie triple system, we can recover Snyder commutation relations, through Jacobson's embedding theorem. In the second subsection, we def\/ine the Weyl map and recover the star product we have introduced.

\subsection{Lie triple system}
In Section \ref{sec:loop}, we have f\/irst introduced the concept of Bol loop. The concept of Bol algebra is the inf\/initesimal version of a smooth Bol loop.
\begin{definition}[left Bol algebra, see \cite{spanish} and references therein]
A left Bol algebra is a vector space $\bb$ equipped with a~bilinear and a~trilinear  product, noted respectively $[\,,\,]_B$ and $\langle\,,\, ,\,\rangle$, satisfying the following properties $X,Y,Z,U,V\in \bb$
\begin{gather}
 \langle X,Y,Z\rangle=-\la Y,X,Z\ra, \qquad [X,Y]_B=-[Y,X]_B,\nn \\
 \la X,Y,Z \ra+\la Y,Z,X\ra+\la Z,X,Y\ra=0, \label{jacobi}\\
  \la U,V,\la Z,X,Y\ra\ra= \la\la U,V,Z\ra, X,Y\ra+\la Z, \la U,V,X\ra, Y\ra+\la Z,X,\la U,V,Y\ra\ra,\label{derivation}\\
  \la U,V, [X,Y]_B\ra=[\la U,V,X\ra,Y]_B+[X,\la U,V,Y\ra]_B+[X,\la U,V,Y\ra]_B\nn\\
  \phantom{\la U,V, [X,Y]_B\ra=}{}
  +\la X,Y,[U,V]\ra+[[U,V],[X,Y]]_B.\nn
\end{gather}
\end{definition}

\eqref{jacobi} encodes the Jacobi identity for the trilinear operation. \eqref{derivation} means that the map $\delta_{X,Y}:\bb\dr \bb$ def\/ined as $\delta_{X,Y}(Z)= \la X,Y,Z \ra$ is a derivation for the ternary operation. Explicitly, if $D$ is a derivation $D$ for the trilinear product one has
\[
D\la X,Y,Z \ra= \la D X,Y,Z \ra+\la X,DY,Z \ra+\la X,Y,DZ \ra. 
\]

\begin{proposition}\label{products}
Given a Bol loop $L$, the  bilinear and trilinear products $\com{\,,\,}_B$, $\la\,,\,,\,\ra $of the Bol algebra are obtained as follows {\rm \cite{nagy}}:
 \begin{gather}
\com{X_1,X_2}_B \equiv -\demi \frac{d^2}{dt_1dt_2} \left((a_1\cdot a_2 )\cdot (a_2\cdot a_1 )\mone\right), \nn \\ 
( X_1,X_2,X_3)  \equiv i\frac{1}{3!} \frac{d^3}{dt_1dt_2dt_3} \left( ( (a_1\cdot a_2 ) \cdot a_3)\cdot  (a_1\cdot (a_2\cdot a_3) )\mone\right), \nn\\
\langle X_1,X_2,X_3\rangle =  2 ( X_1,X_3,X_2)+ \com{\com{X,Y}_B,Z}_B,\label{trilinear bol}
 \end{gather}
 with  $X_i={\frac{d}{dt_i}a_i}\big|_{{t_i=0}}= \frac{d}{dt_i}{e^{it_iX_i}}\big|_{{t_i=0}}$.
 \end{proposition}

 We can identify now the inf\/initesimal structures behind the most well known examples of Bol loops.

\begin{example}\label{lie}
A Lie algebra $\g$ with product $[\,,\,]$ is a  Bol algebra such that
\[
[X,Y]_B\equiv [X,Y], \qquad \la X,Y,Z \ra\equiv 0.
\]
This can be derived from Proposition~\ref{products} using the loop given by a Lie group. We recover that a~Lie algebra is  the inf\/initesimal version of a~Lie group.
\end{example}
\begin{definition}[Mal'tsev algebra, see \cite{spanish} and references therein]
A \textit{Mal'tsev algebra} is a left Bol algebra such that
\[
\la X,Y,Z \ra= [[x,y]_B,z]_B-\frac{1}{3}J(x,y,z), \qquad [J(x,y,z),x]_B= J(x,y,[x,z]_B)
\]
with
\[
J(x,y,z)= [[x,y]_B,z]_B+[[z,u]_B,y]_B+[[y,z]_B,x]_B.
\]
Such algebra can be seen as the inf\/initesimal version of a smooth Moufang loop as it can be checked using Proposition~\ref{products}.
\end{definition}

We introduce now the  relevant example for studying Snyder space-time.

\begin{definition}[Lie triple system, see \cite{spanish, jacobson, kikawa} and references therein]
A\textit{ Lie triple system} $\ell$ is a left Bol algebra such that
\[
 [X,Y]_B=0,\qquad \forall\, x,y\in\ell. 
\]
It is therefore totally determined in terms of the trilinear product which has the following properties
\begin{gather}
\langle X,Y,Z\rangle=-\la Y,X,Z\ra, \nn \\
 \la X,Y,Z \ra+\la Y,Z,X\ra+\la Z,X,Y\ra=0, \label{jacobi 1}\\
 \la U,V,\la Z,X,Y\ra\ra= \la\la U,V,Z\ra, X,Y\ra+\la Z, \la U,V,X\ra, Y\ra+\la Z,X,\la U,V,Y\ra\ra.\nn 
\end{gather}
In fact a direct calculation  from \eqref{trilinear bol} shows that we have
\begin{gather}\label{explicit trilinear product}
\langle X,Y,Z\rangle=2(Z,X,Y)= [[X,Y],Z]= (XY-YX)Z-Z(XY-YX).
\end{gather}
This structure can be seen as the inf\/initesimal version of a smooth K-loop~\cite{kikawa}, once again using Proposition~\ref{products}.
\end{definition}
Lie triple systems have been studied by mathematicians and most of the results found in the context of Lie algebras have been analysed in the Lie triple system context, see~\cite{kikawa, hodge, lister} and references therein.

The property~\eqref{explicit trilinear product} indicates that the trilinear product can be expressed in terms of some Lie algebra bracket. The following theorem makes this statement more precise.

\begin{theorem}[standard imbedding \protect{\cite{jacobson, lister}}]\label{imbedding}
Let $\ell$ a Lie triple system, and for \mbox{$X,Y\in\ell$}, we note $\delta_{X,Y}(Z)$ the linear transformation $Z\dr \la X,Y,Z \ra$ and $\mmm$ the set of all such linear transformations. Then $\mmm$ is a Lie algebra and moreover the  vector space $\ggg\equiv\ell\oplus\mmm$  is a Lie algebra with bracket
\begin{gather*}
\com{X,Y}= \delta_{X,Y}, \qquad \com{A,B}= AB-BA, \\
\com{A,X}= -[X,A]= AX, \qquad \forall \, A,B\in \mmm,  \quad \forall \, X,Y\in \ppp.
\end{gather*}
The map $\sigma:\ggg\dr \ggg$, such that $\sigma(X)=-X$, $\sigma(A)=A$ is an involutive automorphism of $\ggg$.
\end{theorem}

Notice that the bracket appearing in this construction has nothing to do with the Bol brac\-ket~$[\,,\,]_B$.

\begin{remark}
From the property \eqref{derivation}, we notice that $\delta_{X,Y}$ is a derivation. One can restate the previous theorem by saying that one can make ${\rm Der}_\ell\oplus\ell$ a Lie algebra, where ${\rm Der}_\ell$ is the set of (inner) derivations of~$\ell$.
\end{remark}

 \begin{proof}
To prove the Theorem \ref{imbedding}, one has to check that the Lie bracket satisf\/ies the Jacobi identity. Following  the property~\eqref{jacobi 1}, it is clear that we have for example
\[
\com{\com{X,Y},Z }+\com{\com{Z,X},Y }+\com{\com{Y,Z},X }=0.
\]
The relation
\[
\com{\com{A,Y},Z }+\com{\com{Z,A},Y }+\com{\com{Y,Z},A }=0,
\]
states that $A=\delta_{X,Y}$ is a derivation. This relation is shown by straightforward calculations. The other relations follow also by direct calculations.
\end{proof}

Conversely, given a Lie algebra $\ggg$  and an involutive automorphism $\sigma$, one can easily construct a Lie triple system. Indeed, since $\sigma$ is an involution, it has two eigenvalues $-1$, $1$, with respective eigenspaces noted $\ell$, $\mmm$, such that $\ggg= \ell\oplus \mmm$. $\sigma$ is an automorphism, therefore using the compatibility with the product, it is easy to check that $\com{\ell,\ell}\subset \mmm$, $\com{\ell,\mmm}\subset\ell$ and $\com{\mmm,\mmm}\subset\mmm$. The vector space $\ell$ equipped with the trilinear product $\la X,Y,Z \ra=\com{\com{X,Y},Z}$ constructed from the Lie algebra bracket $\la\, ,\, ,\, \ra$ is a Lie triple  system.

This theorem and its converse are the key to understand the nature of Snyder spacetime. The following example shows how to construct a Lie triple system from $\so(p,1)$ and recover Snyder non-commutative structure.

\begin{example}\label{dS triple}
The decomposition of the Lie algebra $\so(p,1)\sim \ell\oplus \so(p-1,1)$ is  given by the involutive automorphism  $\sigma(J_{p\mu})=-J_{p\mu} $ and $\sigma(J_{\mu\nu})=J_{\mu\nu} $ with  $\mu=0,\dots,p-1$  and $J_{\mu\nu}\in\so(p-1,1)$.  The sector $\ell$ is  generated by the ``de Sitter boosts'' $J_{p\mu}$ and is therefore a Lie triple system.  We have following Jacobson's theorem
\[
\langle J_{p\mu}, J_{p\nu}, J_{p\alpha}\rangle= \eta_{\nu\alpha} J_{p\mu}-  \eta_{\mu\alpha} J_{p\nu}.
\]
Conversely, given the Lie triple system $\ell$, we recover the  Snyder  commutation relation by introducing the dimensionful coordinates
\[
X_\mu= \frac{1}{\ka}J_{p\mu},
\]
which satisfy therefore, still thanks to Jacobson's theorem
\[
[X_\mu,X_\nu]\equiv \delta_{X_\mu,X_\nu}=\langle X_\mu,X_\nu, .\rangle=\frac{1}{\ka^2}J_\mn.
\]
\end{example}

This can be  naturally extended to the case $\so(p,q)$.  Coming back to the example we have considered in Section~\ref{sec:field theory}, the smooth K-loop $L\sim \SO(4,1)/\SO(3,1)$ is associated with the Lie triple system of Example \ref{dS triple}. The operator $X_\mu\sim \frac{1}{\ka} J_{p\mu}$ can be viewed as a distribution (with support the identity element) as follows. Given a test function $f\in \cc(L^+)$,
\[
(X_\mu,f)\equiv -2\ka i \frac{d}{d\eta}f \big(e^{i\frac{\eta}{2} J_{p\mu}}\big)\Big|_{\eta=0}.
\]
The bilinear and trilinear products are then constructed in a similar way as in Proposition \ref{products}
\begin{gather*}
(\com{X_1,X_2}_B,f) \equiv i\demi \frac{d^2}{dt_1dt_2} f\left((a_1\cdot a_2 )\cdot (a_2\cdot a_1 )\mone\right), \\ 
(\langle X_1,X_2,X_3\rangle,f)  i\equiv \frac{1}{3} \frac{d^3}{dt_1dt_2dt_3}f \left( ( (a_1\cdot a_2 ) \cdot a_3) \cdot (a_1\cdot (a_2\cdot a_3) )\mone\right).
\end{gather*}
This means in particular that there is two types of bracket one should not confuse. On one hand there is the Bol bracket $[X_\mu,X_\nu]_B=0$ which is trivially zero since we are dealing with a K-loop (and especially the automorphic property).  On the other hand, there is the bracket associated with the imbedding Lie algebra $\com{X_\mu,X_\nu}\equiv \delta_{X_\mu,X_\nu} $ from which the trilinear product is constructed  as in \eqref{explicit trilinear product}. The Snyder commutation relations are encoded in this bracket and \textit{the natural algebraic framework to describe the Snyder space-time given by the Lie triple system structure}.

\subsection{Weyl map}
We have seen in the last section that given a Lie triple system, Jacobson's theorem allows to recover the Snyder commutation relation. It is therefore natural to consider the set of functions on Snyder spacetime as the set of functions on the Lie triple system.

The notion of enveloping algebra $U(\bb)$ has been constructed for  Bol algebras, and in particular it satisf\/ies the Poincar\'e--Birkhof\/f--Witt theorem;  we refer to \cite{spanish} for the details. As a~consequence, such enveloping algebra $U(\ell)$ exists  for a Lie triple system~$\ell$. This enveloping algebra $U(\ell)$ can be interpreted as the  algebra of non-commutative functions $\cc(\ell)$ on Snyder space-time, generated from the coordinates operators~$X_\mu$.

We want to check now that the star product we have introduced in Section~\ref{sec:field theory} is related to the non-commutative structure encoded in the Lie triple system. For this we need to introduce a Weyl map, an algebra isomorphism between the algebra of functions on the Lie triple system  $\cc(\ell)$ and the algebra of functions with the star product $\cc_*(\R^4)$ .
\begin{gather}
\ww: \cc(\ell) \dr  \cc_*\big(\R^4\big), \nn\\
\hat f(X)\equiv \int_{dS^+} [dp]\,  f(p)   e^{ip\cdot X}   \dr  \ww(\hat f)(x) \equiv\int_{dS^+} [dp]\,  f(p)   e^{ip\cdot x},\label{weyl map}
\end{gather}
where $p$ is a momentum coordinate choice on the upper part of the de Sitter space, $[dp]$ the measure on the upper part of the de Sitter space, $e^{ip\cdot X}\in L$  and $f(p)\in \cc(L)$.
For example, for the generator $X_\mu$, we have $\ww(X_\mu)(x)\equiv  x_\mu$.  The Weyl map is an isomorphism of algebra therefore by def\/inition we need to have $\ww(\hat f \cdot \hat h)(x)\equiv (\ww(\hat f)* \ww(\hat h))(x)$. The def\/inition of the Weyl map~\eqref{weyl map} makes sure that the $*$-product we are using is precisely the one def\/ined in  Section~\ref{sec:field theory}.
We have in particular
\[
\ww(X_\mu X_\nu)(x)= x_\mu*x_\nu.
\]
It is then immediate to check that
\[
\ww([X_\mu,X_\nu]_B)= 0=\ww(X_\mu X_\nu- X_\nu X_\mu)= x_\mu*x_\nu- x_\nu x_\mu=0,
\]
in accordance with \eqref{commutator zero}. To consider the mapping of a product of three operators, we need to be careful since we have to take into account the non-associativity.  Indeed, the   position opera\-tor~$X_\mu$   acts by ($*$-)multiplication  of $x_\mu$
\begin{gather*} 
X_\mu \act \hat f(x)\equiv x_\mu * \hat f(x),
\end{gather*}
and we use the natural action on the left, so that when considering the product of operators, we have
\begin{gather*}
(X_\mu X_\nu)\act \hat f(x))\equiv  X_\mu \act (X_\nu\act \hat f(x))= x_\mu  *(x_\nu * \hat f(x)).
\end{gather*}
This is important to keep in mind since we are using now a non-associative structure: order does matter. The product of operators is therefore naturally ordered from the right to the left.  We have then
\[
\ww(X_\mu X_\nu X_\alpha)= x_\mu*(x_\nu*x_\alpha).
\]
When considering the triple product, the commutator does contribute in a non trivial way due to the non-associativity
\begin{gather}
\ww(\langle X_\mu,X_\nu,X_\alpha\rangle) = \ww([ [X_\mu,X_\nu],X_\alpha])= \ww\left(\frac{i}{\ka^2} [J_\mn,X_\alpha]\right)\nn\\
 \phantom{\ww(\langle X_\mu,X_\nu,X_\alpha\rangle}{}
 =  \ww((X_\mu X_\nu- X_\nu X_\mu)X_\alpha- X_\alpha (X_\mu X_\nu- X_\nu X_\mu) )\nn\\
 \phantom{\ww(\langle X_\mu,X_\nu,X_\alpha\rangle}{}
  = x_\mu* (x_\nu*x_\alpha)- x_\nu *(x_\mu*x_\alpha)- x_\alpha *(x_\mu *x_\nu- x_\nu* x_\mu) )\label{intermediate} \\
 \phantom{\ww(\langle X_\mu,X_\nu,X_\alpha\rangle}{}
 =  \frac{i}{\ka^2} [J_\mn,x_\alpha]= \langle x_\mu,x_\nu,x_\alpha\rangle ,\nn
\end{gather}
where we have used \eqref{commutator zero}, so that the last term in \eqref{intermediate} is zero and we have used the result
in~\eqref{star triple product 1}. The Weyl map preserves, as required, the Lie triple system structure.

The Snyder commutation relation $[X_\mu, X_\nu]$ can also be obtained from the $*$-product using its action a function $\hat f(x)$, which we take to be $x_\alpha$  for simplicity \cite{girliv, girliv2}
\[
[X_\mu, X_\nu]\act x_\alpha= \left(X_\mu X_\nu- X_\mu X_\nu\right) \act x_\alpha=x_\mu*(x_\nu * x_\alpha)- x_\nu*(x_\mu * x_\alpha)= \frac{ i}{\ka^2}J_\mn\act x_\alpha.
\]
This shows that the commutator of the  position operators does satisfy
\begin{gather} \label{snyder 3}
[X_\mu,X_\nu]=  i\frac{1}{\ka^2}J_{\mu\nu}.
\end{gather}

\section*{Concluding remarks}
There are two ways to interpret  Snyder commutation relations.
The f\/irst one is to say that this non-commutative space  is actually a subspace of a bigger non-commutative space generated from the coordinates algebra $(X_\alpha,  J_\mn)$. The coordinates $J_\mn$ are interpreted as coordinates describing  extra dimensions.  This is the perspective followed in the Doplicher--Fredenhagen--Roberts model~\cite{dfr} which can be seen then as an abelianization of the Snyder model~\cite{girliv2, carlson} with this interpretation.
The second possibility is to consider space-time as only generated by the~$X_\mu$. In this case, we have to face the issue that the coordinates commutations do not close. We argued in this paper that the solution of this issue is  to consider Snyder space-time given in terms of a Lie triple system, that is instead of the bilinear product~\eqref{snyder 3}, one considers the trilinear product
\[
\langle X_\mu,X_\nu,X_\alpha\rangle\equiv[[X_\mu,X_\nu],X_\alpha]=  \frac{1}{\ka^2}\left(\eta_{\nu\alpha}X_\mu - \eta_{\mu\alpha} X_\nu\right).
\]
In this case, one can construct a meaningful notion of non-commutative algebra of functions using the notion of enveloping algebra for Lie triple system which is dual, using a generalized Fourier transform, to the algebra of functions on the  K-loop $L\sim \SO(p,1)/\SO(p-1,1)$. On the other hand, given the Lie triple system, Jacobson's theorem tells us how to embedd it into a Lie algebra and to recover the Snyder commutation relations, from the trilinear product
\[
[X_\mu,X_\nu]\equiv \delta_{X_\mu,X_\nu}=\langle X_\mu,X_\nu, .\rangle=\frac{1}{\ka^2}J_\mn.
\]
Lie triple systems  provide   therefore a new type of non-commutative geometry. This non-commutative geometry can be seen as f\/lat since one can identify a quantum group, a deformation of the the Poincar\'e group, which acts on this space in a consistent way. We  provided here the def\/inition of this new quantum group, which we called \textit{Snyder quantum group}, using the structure of  K-loop. This latter can be be interpreted as momentum space as in Snyder's initial idea. The K-loop structure provides  a rule to add momenta, an addition which is non-commutative and non-associative, just like the speed addition in Special Relativity.  We have also  presented  how we can construct a scalar f\/ield action on a Lie triple system using a star product realization  and how this  action is invariant under the Snyder quantum group, encoding the analogue of the f\/lat symmetries.

This work opens up a number of questions which are of interests for either  mathematicians or  theoretical physicists.
\begin{itemize}\itemsep=0pt
\item The Snyder quantum group needs to be studied in details:  the  analysis of the representation theory should  be performed. It would be interesting to check if the weak notion of commutativity \eqref{gyrocom} is related to a notion of braiding.
\item  The classif\/ication of the bi-covariant dif\/ferential calculus \cite{sitarz} for Snyder space-time should be performed to make the analysis of the conserved currents~\cite{laurent-kowalski}.
\item Now that we have a well def\/ined classical action for a scalar f\/ield, it would be interesting to check if this non-commutative space fullf\/ils Snyder's hope, that is whether a quantum f\/ield theory living in this space has no UV divergences.
\end{itemize}

\appendix
\section{Snyder sum}
In this appendix we present the calculations to determine the modif\/ied sum of momenta \eqref{snyder sum1}, \eqref{snyder sum2}  in Snyder's coordinates. We calculate the product of de Sitter boosts $a_i=\cosh\frac{\eta_i}{2}\, {\rm Id} + 2i\sinh \frac{\eta_i}{2}  B^\mu J_{4\mu}$
\[
a_1a_2=a \Lambda R,
\]
where $\Lambda= \cosh\frac{\alpha}{2}\, {\rm Id} + 2i\sinh \frac{\alpha}{2}\, b^iJ_{0i}$ is a boost and $R= \cos\thet\, {\rm Id} + i \sin\thet \, r^i J_i$ is a rotation. The term $a$ will encode the sum of momenta and $\Lambda R$ is the Lorentz precession, the generalization of the Thomas precession to the 4d Lorentzian case \cite{dsr etfl}
\begin{alignat}{3}
& \cosh\etaa\cosh\etab- \sinh\etaa\sinh\etab B_1\cdot B_2  =  \cosh\et\, \cosh \alp \cos\thet \quad &&  (\one), & \label{id}\\
& \sinh\etaa\sinh\etab (\vec B_1\wedge \vec B_2)^i =  \cosh \et \, \cosh\alp \cos\thet \, r^i \quad && (R_i),& \label{R}\\
& \sinh\etaa\sinh\etab (B_{1}^0B_2^i-B_1^iB_2^0) = \cosh\et \sinh\alp\left(\cos\thet b^i+\sin\thet (\vec b\wedge \vec r)^i\right) \  &&  (K), & \label{K}\\
& \cosh\etaa\sinh\etab\, B_2^0+\cosh\etab\sinh\etaa\, B_1^0 =  \cosh\alp \sinh\et \cos\thet \, B^0  &&& \nn\\
&  \qquad\quad{} +\sinh\et\sinh\alp\left(\cos\thet \vec B\cdot \vec b+\sin\thet \epsilon_{jki}r^iB^jb^k\right)\quad && (J_{40}), & \label{J40}\\
& \cosh\etaa\sinh\etab\, B_2^i+\cosh\etab\sinh\etaa\, B_1^i &&& \nn \\
& \qquad {}= \cosh\alp\sinh\et\left(\cos\thet B^i-\sin\thet\epsilon_{mk}^ir^mB^k\right) &&& \nn\\
 & \qquad\quad{} + B^0\sinh\alp\sinh\et\left(\cos\thet b^i-\sin\thet b_k r_j \epsilon_{kj}^i\right) \quad && (J_{4i}). & \label{J4i}
\end{alignat}
The other equations state that we have $\vec r\cdot \vec b=0$ and $\vec r\cdot \vec B=0$. This all together allows to reconstruct the sum of momenta given the choice of coordinates $ k_\mu= \ka \, f(\eta)B_\mu$. we are going to determine this sum in the Snyder's choice, i.e.\ $p_\mu=\ka\tanh\eta\, B_\mu$. With this choice, we have in particular $ \cosh\eta\equiv \Gamma=\frac{1}{\sqrt{1+\frac{p^2}{\ka^2}}}$.
Considering the ratio of the equation \eqref{R} with \eqref{id}, together with  $\tanh\et=\frac{\tanh\eta}{1+\Gamma\mone}$, we have
\[
\tan\thet\,\vec r =  \frac{\vec p_1\wedge \vec p_2}{\ka^2(1+\Gamma_1\mone)(1+\Gamma_2\mone)-p_1\cdot p_2}
\ \Leftrightarrow\ \vec \rho= \vec A(p_1,p_2), 
\]
with $\vec \rho= \tan\thet\,\vec r$. Since ${\vec r}\,{}^2=1$, we have in particular that
\begin{gather}\label{tan}
\tan^2\thet= |\vec A(p_1,p_2)|^2.
\end{gather}
Considering the ratio of the equation \eqref{K} with \eqref{id}, we obtain
\begin{gather*} 
\tanh\alp \left(\vec b-\tan\thet \vec r\wedge\vec b\right) =  \frac{p_1^0\vec p_2-p_2^0\vec p_1  }{\ka^2(1+\Gamma_1\mone)(1+\Gamma_2\mone)-p_1\cdot p_2}\\
\qquad{} \Leftrightarrow \ \tanh\alp \left(\one-\tan\thet \vec r\wedge\right)(\vec b) =  \frac{p_1^0\vec p_2-p_2^0\vec p_1  }{\ka^2(1+\Gamma_1\mone)(1+\Gamma_2\mone)-p_1\cdot p_2}\\
\qquad \Leftrightarrow\ \vec F(\vec \beta) = \vec C(p_1,p_2),
\end{gather*}
with $\vec \beta= \tanh\alp\, \vec b$. The map $\vec F$ is a linear map and bearing in mind that $\vec r\cdot b=0$, its inverse~$\vec F\mone$ is given by
\[
\vec F\mone= \cos^2\thet \left(\one + \tan\thet\, \vec r\wedge\right)=\cos^2\thet \, \vec D,
\]
and is also a linear map.  We can therefore def\/ine $\vec \beta= \tanh\alp\, \vec b$ in terms of $p_1$ and $p_2$ since
\begin{gather*} 
\vec \beta =  \cos^2\thet \,\vec D(\vec C(p_1,p_2))
 = \cos^2\thet (\one +  \vec \rho\wedge)\left(\frac{p_1^0\vec p_2-p_2^0\vec p_1  }{\ka^2(1+\Gamma_1\mone)(1+\Gamma_2\mone)-p_1\cdot p_2}\right).
\end{gather*}
Now since $\cos^2\thet = \big({1+\tan^2\thet}\big)\mone$ and $\tan^2\thet $ is given in \eqref{tan}, we can determine $\cos^2\thet$ in terms of $A(p_1,p_2)$ and hence $\vec \beta$ in terms of $p_1$ and $p_2$.

Considering the ratio of the equation \eqref{J4i} with \eqref{id}, we obtain
\begin{gather*}
\ka\left(\frac{(1+\Gamma_2\mone)\,\vec p_1+ (1+\Gamma_1\mone)\,\vec p_2}{\ka^2(1+\Gamma_1\mone)(1+\Gamma_2\mone)-p_1\cdot p_2}\right) =  \tanh\et (\vec B- \vec \rho\wedge \vec B) +\tanh\et\,  B^0 ( \vec \beta+ \vec \rho \wedge \vec \beta),\\
\vec E(p_1,p_2) =  \tanh\et \vec F (\vec B)+ \tanh\et\,  B^0 \vec D(\vec \beta).
\end{gather*}
We can use the inverse map $\vec F\mone= \cos^2\thet \,\vec D$ to determine $\vec B$ in terms of $B_0$, $p_1$, $p_2$ and $\eta$, since we have that $\vec r\cdot \vec B=0$,
\begin{gather}
 \tanh\et \vec B  =  -\tanh\et\,  B^0 \,\cos^2\thet \,\vec D(\vec D(\vec \beta))+ \cos^2\thet\,\vec D(\vec E(p_1,p_2)) \nn\\
\qquad \Leftrightarrow \ \tanh\et \vec B  =   -\tanh\et\,  B^0  \, \vec V + \vec W, \label{Bi}
\end{gather}
where $\vec V\equiv\cos^2\thet \,\vec D (\vec D(\vec \beta) )$ and $\vec W\equiv\cos^2\thet\,\vec D (\vec E(p_1,p_2) )$ are vectors totally determined in terms of~$p_1$ and~$p_2$.

Considering the ratio of the equation \eqref{J40} with \eqref{id}, we obtain
\begin{gather*}
\ka\left(\frac{(1+\Gamma_2\mone)\,p_1^0+ (1+\Gamma_1\mone)\,p_2^0}{\ka^2(1+\Gamma_1\mone)(1+\Gamma_2\mone)-p_1\cdot p_2}\right) =  \tanh\et\, \big(B^0+ (\vec B\cdot \vec \beta+\vec B\cdot (\vec \rho \wedge \vec \beta))\big) \\
\qquad{}  \Leftrightarrow \ I(p_1,p_2) =   \tanh\et\, \big(B^0+ \vec B \cdot \vec D(\vec \beta) \big).
\end{gather*}
We can used now the value for $\vec B$ identif\/ied in \eqref{Bi}
\begin{gather*}
\tanh\et\, B^0+ \left(-\tanh\et\,  B^0  \, \vec V + \vec W\right) \cdot \vec D(\vec \beta)  = I(p_1,p_2)\\
\qquad {} \Leftrightarrow \ \tanh\et\, B^0 \big(1-  \vec V  \cdot \vec D(\vec \beta)\big) =  - \vec W\cdot \vec D(\vec \beta) +I(p_1,p_2) \\
\qquad {} \Leftrightarrow \ \tanh\et\, B^0  = \frac{ - \vec W\cdot \vec D(\vec \beta) +I(p_1,p_2)}{1-  \vec V  \cdot \vec D(\vec \beta)}.
\end{gather*}
We can then plug the value of $B_0$ into \eqref{Bi} to  obtain $B_i$ in terms of the $p_i$
\[
\tanh\et \vec B = \frac{ - \vec W\cdot \vec D(\vec \beta) +I(p_1,p_2)}{1-  \vec V  \cdot \vec D(\vec \beta)} \, \vec V + \vec W.
\]
The f\/inal step to recover the sum is to recall that  $\tanh\et= \frac{\tanh\eta}{1+\Gamma\mone}$ and  $\Gamma$ can  be expressed  in terms of $\Gamma_i$
\[
\Gamma= \Gamma_1\Gamma_2\left(1-\frac{p_1\cdot p_2}{\ka^2}\right).
\]
 Hence we have that
\begin{gather*}
\vec p_{\rm tot}=  \overrightarrow{p_1\oplus  p_2} =  \left(1+  \Gamma_1\mone\Gamma_2\mone\left(1-\frac{p_1\cdot p_2}{\ka^2}\right)\mone\right) \left(\frac{ - \vec W\cdot \vec D(\vec \beta) +I(p_1,p_2)}{1-  \vec V  \cdot \vec D(\vec \beta)} \, \vec V + \vec W\right),\\
p^0_{\rm tot}= (p_1\oplus p_2)^0 =  \left(1+  \Gamma_1\mone\Gamma_2\mone\left(1-\frac{p_1\cdot p_2}{\ka^2}\right)\mone\right)\frac{ - \vec W\cdot \vec D(\vec \beta) +I(p_1,p_2)}{1-  \vec V  \cdot \vec D(\vec \beta)}.
\end{gather*}

\pdfbookmark[1]{References}{ref}
\LastPageEnding

\end{document}